\documentclass[aps,pra,twocolumn,superscriptaddress,amsmath,showpacs,amssymb]{revtex4} 

\usepackage{graphicx}
\usepackage{amsthm,amsmath}
\usepackage{color}

\newcommand{\be}{\begin{eqnarray}}
\newcommand{\ee}{\end{eqnarray}}

\newcommand{\nn}{\nonumber}

\newcommand{\bo}{\boldsymbol}
\newcommand{\lb}{\label}

\newcounter{ichi}
\newcounter{ni}
\newcounter{san}
\newcounter{yon}
\newcounter{go}
\newcounter{roku}
\newcounter{nana}
\newcounter{hati}
\newcounter{kyu}
\begin{document}


\title{Nontrivial Haldane phase of an atomic two-component Fermi gas trapped 
in a 1d optical lattice}

\author{Keita Kobayashi}
\affiliation{CCSE, Japan Atomic Energy Agency, 5-1-5 Kashiwanoha, Kashiwa, Chiba 277-8587, Japan}
\affiliation{CREST(JST), 4-1-8 Honcho, Kawaguchi, Saitama 332-0012,
Japan}
\author{Masahiko Okumura}
\affiliation{CCSE, Japan Atomic Energy Agency, 5-1-5 Kashiwanoha, Kashiwa, Chiba 277-8587, Japan}
\affiliation{CREST(JST), 4-1-8 Honcho, Kawaguchi, Saitama 332-0012, Japan}
\affiliation{Computational Condensed Matter Physics Laboratory, RIKEN, Wako, Saitama 351-0198, Japan}
\affiliation{Computational Materials Science Research Team, RIKEN AICS, Kobe, Hyogo 650-0047, Japan}
\author{Yukihiro Ota}
\affiliation{Advanced Research Institute, RIKEN
2-1 Hirosawa, Wako-shi, Saitama 351-0198, JAPAN}
\author{Susumu Yamada}
\affiliation{CCSE, Japan Atomic Energy Agency, 5-1-5 Kashiwanoha, Kashiwa, Chiba 277-8587, Japan}
\affiliation{CREST(JST), 4-1-8 Honcho, Kawaguchi, Saitama 332-0012, Japan}
\affiliation{Computational Materials Science Research Team, RIKEN AICS, Kobe, Hyogo 650-0047, Japan}
\author{Masahiko Machida}
\affiliation{CCSE, Japan Atomic Energy Agency, 5-1-5 Kashiwanoha, Kashiwa, Chiba 277-8587, Japan}
\affiliation{CREST(JST), 4-1-8 Honcho, Kawaguchi, Saitama 332-0012, Japan}
\affiliation{Computational Materials Science Research Team, RIKEN AICS, Kobe, Hyogo 650-0047, Japan}

\date{\today}

\begin{abstract}  

We propose how to create a non-trivial Haldane phase in atomic
 two-component Fermi-gas loaded on one-dimensional (1-D) optical lattice
 with trap potential. 
The Haldane phase is naturally formed on $p$-band Mott core in a wide range of the strong
 on-site repulsive interaction. 
The present proposal is composed of two steps, one of which is
 theoretical derivation of an effective 1-D $S=1$ interacting-chain
 model from the original tight-binding Hamiltonian handling the two
 $p$-orbitals, and the other of which is numerical demonstration
 employing the density-matrix renormalization-group for the formation of the Haldane phase on
 $p$-band Mott core and its associated features in the original tight-binding model with
 the harmonic trap potential. 

\end{abstract}

\maketitle


Ultra-cold Fermi gas loaded on optical lattice (FGOL) is one of the most
fascinating systems in modern physics \cite{FGOL}. 
Its controllability is far beyond our past image and experience in
condensed matters. 
For instance, the two-body interaction between Fermi atoms is widely
variable from strongly attractive to repulsive. 
Moreover, filling and population imbalance are freely tunable, and
lattice geometry is highly flexible. 
Thus, FGOL is regarded as an outstanding simulator for
strongly-correlated electronic materials, in which a great number of
controversial issues like high-temperature superconducting mechanism in
cuprate superconductors still remain unsolved. 

Very recently, there have been intense theoretical and experimental
interests in extending the energy band produced by the optical lattice
(OL) potential from the ground single one to higher multiple-ones by
utilizing multi-degenerate higher orbitals \cite{p-band,p-band2}.
Such multiple orbital degeneracy allows us to study orbital degrees of
freedom in addition to charge and spin ones.  
Then, FGOL becomes a more realistic and rich simulator, in which orbital
ordering and high-spin correlation are accessible subjects like real
solid-state matters including transition metals and other heavy elements
\cite{spin3/2,orbital order}. 
In this paper, we study one of the most fundamental issues associated
with the multiple band degeneracy. 
The present target is formation of a gapped quantum phase using double
$p$-orbital degeneracy under a one-dimensional(1-D) system, as shown in
Fig.1. 
We show that an effective low-energy Hamiltonian is given by
the $S=1$ Heisenberg model and the gapped Haldane phase\,\cite{Haldane}
is widely sustained by the harmonic trap potential to confine Fermi
atoms. 
The fertility of the Haldane phase, i.e., gapful
spin-excitation, nonlocal string order, and spin-1/2 edge magnetization
induction \cite{AKLT,string order,Miyashita,spin1-NMR} can be
systematically explored because of the wide controllability in FGOL's.   
\begin{figure}[h]
\begin{center}
\includegraphics[width=1.0\linewidth]{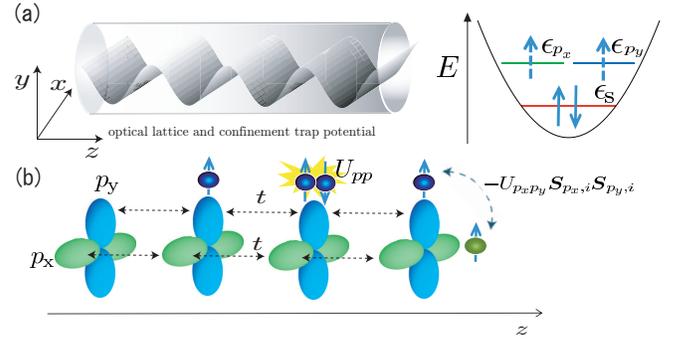}
\end{center}
\caption{(Color Online) Schematic figures for a 1-D Fermi gas. 
(a) An optical lattice (OL) potential along $z$-direction and a
 cylindrically-symmetric vertical trap-potential inside $xy$-plane. 
The latter causes to form discrete levels as depicted at
 the right panel, in which the lowest level is $s$-orbital while higher
 two degenerate ones are $p_{x,y}$-obitals. 
The Haldane phase is formed under fully-filled $s$- and partially-filled
 $p$-orbitals (see text). 
(b) $p_{x,y}$-orbitals formed inside the vertical trap and their
 $p$-bands creatd by the OL. 
When $p$-bands are active, this system is
 described by a multi-band Hubbard chain with the on-site intra-orbital
 interaction ($U_{pp}$) and inter-orbital one ($U_{p_x p_y}$) (see
 Eq.(\ref{eq:p1d}) and text).
}
\label{schematic fig}
\end{figure}

Feasibility studies to create the Haldane phases in atomic gas OL's
have been made on spin 1 bosonic gases loaded on 1-D OL with one atom per site\,\cite{S=1 boson},  
two-component 1-D FGOL utilizing 1st and 2nd Bloch-bands with Feshbach resonance\,\cite{band-Haldane}, and 
ultra-cold fermionic $^{171}{\rm Yb}$ in 1-D OL\,\cite{Hund}\,.
The Haldane insulator, in which a charge gap opens
according to Haldane's conjecture\,\cite{Haldane}, 
was also predicted for bosonic gases 
with dipole interaction\,\cite{Haldane boson1,Haldane boson1.5,Haldane boson3} and multi-component 1-D FGOL 
\cite{Haldane fermi}.
In this paper, through the derivation of an effective $S=1$
interacting-chain model and the numerical simulations of a multi-band
Hubbard model, we show that the Haldane phase is naturally formed in a
two-component 1-D FGOL with the emergence of $p$-band Mott core.  
We also reveal features of the Haldane phase in trapped systems. 


We start with the following Hamiltonian describing a 1-D elongated FGOL
schematically depicted in Fig.1, 
\be 
H
&=&
\sum_{\sigma=\uparrow,\downarrow}\int d\bo{x}
\left[
\psi_{\sigma}^{\dagger}h_{0}\psi_{\sigma}
+
\frac{g}{2}
\psi_{\sigma}^{\dagger}\psi_{\bar{\sigma}}^{\dagger}
\psi_{\bar{\sigma}}\psi_{\sigma}
\right]\,,
\ee
with 
$
h_{0}
=
(-\hbar^2/2m)\nabla^2
+V_{\rm ver}+V_{\rm opt} 
$ 
and the coupling constant of the two-body interaction $g$. 
The cylindrically-symmetric vertical trap (on $xy$-plane) and the OL
potential (along $z$-axis) are, respectively, $V_{\rm ver}$ and $V_{\rm opt}$.  

A multi-band Hubbard-type model is derived from Eq.(1) using the
expansion 
$
\psi_{\sigma}=\sum_{\alpha,\beta}\sum_{i}
c_{\alpha,\beta,\sigma,i}u_{\alpha}w_{\beta,i}\,,
$
where $u_{\alpha}$ and $w_{\beta,i}$ are a wavefunction
associated with the eigensystem of 
\(
\left[(-\hbar^2/2m)\nabla_{\bot}^2+V_{\rm
ver}\right]u_{\alpha}=\epsilon_{\alpha}u_{\alpha}
\) and a Wannier function formed by the OL potential. 
The indices $\alpha$ and $\beta$ represent discrete levels caused by the
trap potential $V_{\rm ver}$ and Bloch-band by the OL potential,
respectively.  
Here, $V_{\rm ver}$ is not so tight that the $2$nd $p$-orbital
energy-level does not exceed over that of $2$nd Bloch-band and the $2$nd
level is partially filled, as shown in Fig.1(b). 
Hereafter, ignoring the contribution from higher Bloch-bands, we drop
the index $\beta$\,. 
Then, including the $2$nd level corresponding to $p_{x(y)}$-orbital and
taking the tight-binding approximation, we obtain a multi-band 1-D
Hubbard Hamiltonian\,\cite{supplementary}
\be
&&H=-\sum_{\alpha,\sigma}
\sum_{<i,j>}t c_{\alpha,\sigma,i}^{\dagger}c_{\alpha,\sigma,j}
+\sum_{\alpha,\sigma,i}\epsilon_{\alpha}
n_{\alpha,\sigma,i}\nn \\
&&+\sum_{i}\bigg{[}
\sum_{\alpha}U_{\alpha\alpha}
n_{\alpha,\uparrow,i}n_{\alpha,\downarrow,i}
-\sum_{\alpha\neq \alpha'}
U_{\alpha\alpha'}\Big{\{}\bo{S}_{\alpha,i}\cdot\bo{S}_{\alpha',i} \nn \\
&&-c_{\alpha,\uparrow,i}^{\dagger}c_{\alpha,\downarrow,i}^{\dagger}
c_{\alpha',\downarrow,i}c_{\alpha',\uparrow,i}
-\sum_{\sigma,\sigma'}\frac{1}{4}n_{\alpha,\sigma,i}n_{\alpha'\sigma',i}
\Big{\}}
\bigg{]}\lb{eq:p1d}  \,,
\ee
where 
\(
n_{\alpha,\sigma,i}
(\equiv c_{\alpha,\sigma,i}^\dagger c_{\alpha,\sigma,i})
\) 
is the on-site number-density operator of $\alpha$-orbital with
pseudo-spin $\sigma$ and $\bo{S}_{\alpha,i}=\frac{1}{2}\sum_{\sigma,\sigma'}
c_{\alpha,i,\sigma}^{\dagger}\bo{\tau}_{\sigma,\sigma'}
c_{\alpha,i,\sigma'}$ with Pauli matrices $\bo{\tau}$ is the local spin-$1/2$ operator\,. 
The summation of $\alpha$ is taken from the ground $s$-orbital to the
$2$nd degenerate $p$-orbitals ($p_{x}$ and $p_{y}$) as shown in
Fig.1(c).  
$t$ and $U_{\alpha\alpha'}$ are the hopping and on-site interaction
energy integrals defined as 
$t=-\int dz\,w_{i+1}
\left(\frac{-\hbar^2}{2m}\frac{\partial^2}{\partial z^2}+V_{\rm opt}\right)
w_{i}$ and
$U_{\alpha\alpha'}=g\int d\bo{x}w_{i}^{4}u_{\alpha}^{2}
u_{\alpha'}^{2}$\,, respectively. 
The in-plane cylindrical symmetry of $u_{\alpha}(\bo{x}_{\bot})$
gives the relations $U_{sp_x}=U_{sp_y} (\equiv U_{sp})$ and
$U_{p_xp_x}=U_{p_yp_y} (\equiv U_{pp})$. 
A significant feature in Eq.(\ref{eq:p1d}) is 
the existence of Hund's like terms 
$-U_{\alpha\alpha'}\bo{S}_{\alpha,i}\cdot\bo{S}_{\alpha',i}$\,, 
which captures the so-called orbital physics. 
Similar Hamiltonian to Eq.(\ref{eq:p1d}) may also appear in ultra-cold fermionic
$^{171}{\rm Yb}$ with $^{1}S_0$- and $^{3}P_0$-states\,\cite{Yb,Hund}\,. 

Now, we derive a $S=1$ Heisenberg chain from Eq.(\ref{eq:p1d}) in the strong coupling limit
($U_{\alpha\alpha'} \gg t$). 
We consider a case when the $1$st orbital is fully occupied and
$2$nd ones are half-filled. 
Now, this system seem to be similar to two Heisenberg chains 
coupled by ferromagnetic exchange interaction, 
which can produce the Haldane phase \cite{Hei}. 
In fact, the second order perturbation scheme\,\cite{BW} leads to 
\(
H_{J}=J_{\rm ex}\sum_{<i,j>}
\bo{S}_{i}\cdot\bo{S}_{j} 
\), 
where $J_{\rm ex}=2t^{2}/(U_{pp}+U_{p_xp_y})$ 
and the local spin-$1$ operator 
\( 
\bo{S}_{i}
=
\frac{1}{2}
\sum_{\alpha=p_x,p_y}
\sum_{\sigma,\sigma'}
c_{\alpha,\sigma',i}^{\dagger}
\bo{\tau}_{\sigma'\sigma}
c_{\alpha,\sigma,i}
\)\,\cite{supplementary}. 
Hence, a gapped Haldane phase can emerge for large $U_{pp}$ 
if the 2nd $p$-levels are half-filled. 
In FGOL experiments, this requirement is achievable in a central Mott
core formed by interplay between a trap potential and large $U_{pp}$. 

\begin{figure}[htbp]
\begin{center}
\includegraphics[width=1.0\linewidth]{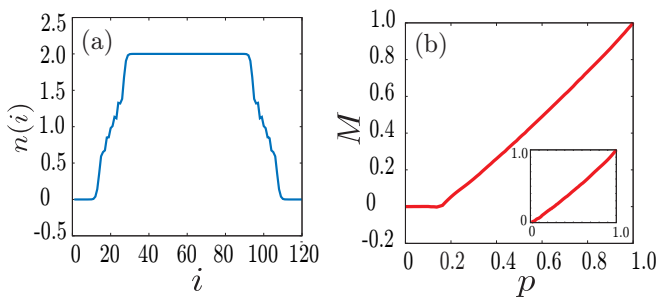}
\end{center}
\caption{(Color Online) (a) A typical spatial distribution profile of
 the particle density 
$n(i) =\sum_{\alpha=p_x,p_y}\sum_{\sigma}n_{\alpha,\sigma,i}$ for
 balanced 160 fermions $(80\uparrow,80\downarrow)$  
with the interaction constants 
$U_{pp}/t=15$\,, $U_{p_x p_y}=\frac{4}{9}U_{pp}$\,, 
and the trap potential strength $V/t=10$\,. 
(b) The Mott-core polarization $M$ vs. the population imbalance ratio
 $p$. 
The employed parameters are the same parameter as in (a) except for $p$. 
For comparison, the inset is the case of $U_{pp}/t=15$\ and 
$U_{p_x p_y}/t=0$,  which is equivalent with two independent $S=1/2$
 interacting chains.}
\label{susceptibility}
\end{figure}

Next, let us turn to the numerical demonstration of the Haldane phase 
by the density-matrix renormalization-group (DMRG) method
\cite{DMRG,DMRG2}. 
Typical atomic-gas experiments employ harmonic trap potential in all
directions to prevent escape of atoms. 
We add the harmonic trap along $z$-axis, 
$V_{\rm ho}(i)=V[2/(L-1)]^{2}[i-(L+1)/2]^{2}$ to our system, 
as well as the aforementioned vertical trap $V_{\rm ver}$. 
Here $L$ means the total number of sites. 
The total Hamiltonian is 
$H+\sum_{\alpha,\sigma,i}V_{\rm ho}(i)n_{\alpha,\sigma,i}$. 
Again, we drop off the terms in the Hamiltonian relevant with the
$1$st orbital\,\cite{supplementary}. 
We simulate the model handling only $2$nd $p$-orbitals\,\cite{condition}.
In the 2-band Hubbard model, the on-site intra-orbital interaction $U_{pp}$ and
inter-orbital interaction $U_{p_xp_y}$ are mutually connected via
$U_{p_xp_y}=\frac{4}{9}U_{pp}$. 
We note that 
$U_{\alpha\alpha'}=g\int d\bo{x}w_i^4 u_\alpha^2u_{\alpha'}^2$\,
\cite{harmonic}.  
Now, we present numerical results. 
We obtain a particle-density distribution 
$n(i)=\sum_{\alpha=p_x,p_y,\sigma}n_{\alpha,\sigma,i}$ 
and a spin-density distribution 
\(
m(i)=\sum_{\alpha=p_x,p_y}
(n_{\alpha,\uparrow,i}-n_{\alpha,\downarrow,i})
\) using DMRG simulations. 
Figure \ref{susceptibility}(a) shows $n(i)$ for $U_{pp}/t=15$ and $V/t=10$\,. 
Owing to $V_{\rm ho}(i)$, $p$-band Mott plateau is
formed in the trap center and surrounded by regions with
trap-gradient-dependent filling (below half). 
Varying an input imbalance ratio $p\equiv \sum_{i}m_{i}/n_{i}$, which is
initially set in experiments, we measure the normalized polarization on
the Mott core\,\cite{Okumura}, 
$M=\sum_{i\in {\rm Mott}}m_{i}/n_{i}$, 
as seen in Fig.\,\ref{susceptibility}(b). 
For comparison, we also show in the inset the result for $U_{p_{x}p_{y}}=0$,
which corresponds to two identical 1-D Hubbard chains without coupling. 
A plateau-like behavior occurs on the Mott-core polarization up to a
critical imbalance ratio $p_{\rm c}(\simeq 0.17)$ as expected in the
presence of the gap, while for $U_{p_xp_y}=0$ the Mott-core is smoothly
magnetized. 
A gapful phase for the spin flip excitation is found in the 
$p$-band Mott-core region. 
Thus, one of the key features of the Haldane phase is shown. 

\begin{figure}[htbp]
\begin{center}
\includegraphics[width=1.0\linewidth]{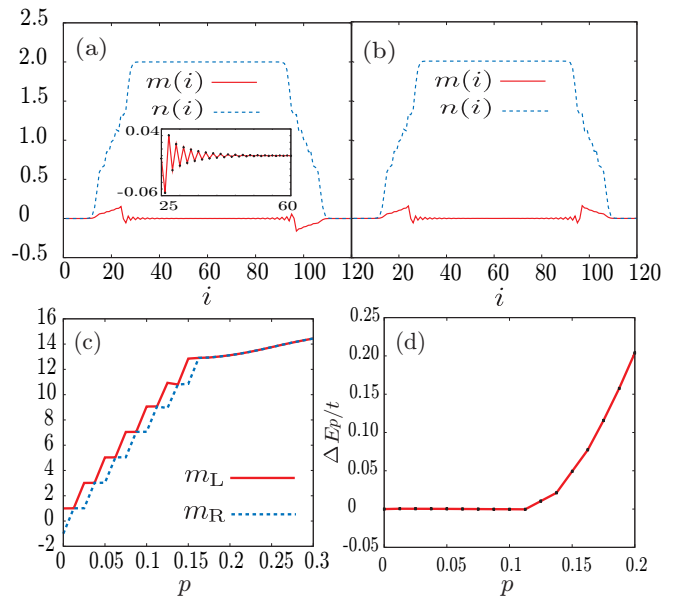}
\end{center}
\caption{(Color Online) Distribution profiles of particle density
 (dashed blue line) $n(i)$  and spin density (solid red line) 
$m(i) =\sum_{\alpha=p_x,p_y}(n_{\alpha,\uparrow,i}-n_{\alpha,\downarrow,i})$ 
 for spin imbalance (a) $p=0$\,, $(S_{\rm tot}^{(z)}=0)$ and 
(b) $p=0.0125$\,, $(S_{\rm tot}^{(z)}=1)$\,. 
The inset of (a) is a focus on the spin density 
around the edge of the left Mott core. 
(c) The polarization in the left (right) surrounding phase  
$m_{\rm L(R)}\equiv \sum_{i\in {\rm left}({\rm right})}m(i)$ 
versus the population imbalance ratio $p$. 
(d) The spin gap $\Delta E_{p}\equiv E_{p}-E_{p=0}$ versus  
the population imbalance ratio $p$\,. 
All other physical parameters are the same as in
 Fig.\ref{susceptibility}(a). 
}
\label{edge_spin1/2}
\end{figure}

Let us more carefully examine the trap center and the outer regions
for a balanced case $p=0$ 
($S_{\rm tot}^{(z)}(\equiv\sum_i S_i^{(z)})=0$) 
and a slightly-polarized case $p=0.0125$ ($S_{\rm tot}^{(z)}=1$). 
First, we focus on the spin-density distribution profiles
on the $p$-band Mott core. 
Figure \ref{edge_spin1/2} shows that staggered magnetization
structures emerge at the edges of the $p$-band Mott core and
exponentially decay toward the trap center. 
This structure is the same as the typical one observed at open boundary edges
in $S=1$ Heisenberg chains\,\cite{Miyashita,spin1-NMR} 
and is known to be induced by $S=1/2$ fraction left at open boundary edges. 
In the present case, the $p$-band Mott core is not sharply
terminated by the open boundary condition. 
The edge magnetization amplitude in the induced staggered structures is
sizeably reduced, compared to the results of $S=1$ Heisenberg chain in
the open boundary case\,\cite{Miyashita}. 
This difference comes from coupling between $S=1/2$ fraction left at
the edges of the $p$-band Mott core and a ferromagnetic metallic phase
in the outer regions. 
The ferromagnetic property in the outer regions is confirmed by
calculating spin-gap energy $\Delta E_{p}=E_{p}-E_{p=0}$, polarization
on outer regions, and the Mott-core polarization\,\cite{supplementary}. 
Here, $E_{p}$ is the ground-state energy for the population imbalance
ratio $p$. 
This ferromagnetism is also consistent with theoretical prediction for
a uniform hole-doped two-degenerate-band system with strong Hund's
coupling\,\cite{double}. 
Next, let us examine the outer regions to understand coupling between
the ferromagnetic metal and the $S=1/2$ fraction at the edge of the Mott
core. 
As for $S_{\rm tot}^{(z)}=0$, as shown in Fig.\,\ref{edge_spin1/2}(a), the
spin-density profile is anti-centrosymmetric. 
We calculate the integral of $m(i)$ over the left (right) outer 
region, $m_{\rm L(R)}\equiv\sum_{i\in {\rm left (right)}}m(i)$. 
Then, we find that the left and right regions, respectively,
are polarized as one up-spin ($m_{\rm L} \simeq 1$) and one down-spin
($m_{\rm R} \simeq -1$). 
For $S_{\rm tot}^{(z)}=1$, as shown in Fig.\ref{edge_spin1/2}(b), both
regions are polarized as one up-spin 
($m_{\rm L}=m_{\rm R}\simeq 1$). 
We always find similar features for various $p$. 
The polarization difference becomes either 
$|m_{\rm L}-m_{\rm R}|\simeq 0$ or  
$|m_{\rm L}-m_{\rm R}|\simeq 2$ as shown in
Fig.\,\ref{edge_spin1/2}(c). 
These results mean that $S=1/2$ fractions left at the edges of the
$p$-band Mott core may spread over the outer regions. 
We further characterize the ferromagnetic metal on the outer regions. 
We compare the spin-gap energy with Mott-core polarization. 
The ground-state degeneracy (i.e., $\Delta E_{p}=0$) exists up to
$p\simeq 0.11$, as seen in Fig.\,\ref{edge_spin1/2}(d), while
plateau-like behavior in Fig.\,\ref{susceptibility}(b) still continues
up to $p\simeq 0.17$. 
This result implies that all spin degrees of freedom in the system
accumulate on the ferromagnetic metal before $p\simeq 0.11$\,. 
Therefore, the ferromagnetic metal on the outer regions corresponds to
an edge state in uniform $S=1$ Heisenberg model. 

Here, we mention the case of small $U_{pp}/t$. 
The Haldane phase may survive in small $U$ region as discussed in the case of 
multi-component 1-D FGOL\,\cite{Haldane fermi}. 
We emphasize that, for a wide-range of the interaction strength and 
the trap potential strength $V/t$, 
the gapped Mott core more or less appears 
and the Haldane phase is realized in the central core
region\,\cite{supplementary}. 

\begin{figure}[h]
\begin{center}
\includegraphics[width=1.0\linewidth]{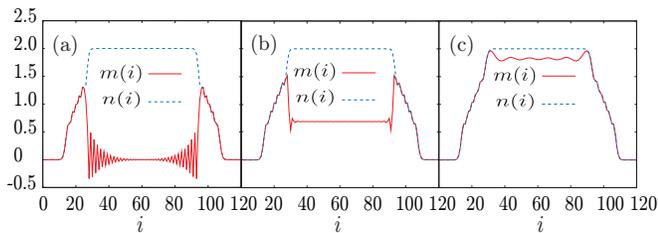}
\end{center}
\caption{(Color Online) Particle density (dashed blue line) and 
spin density (solid red line) distribution profiles for 
population imbalance ratios (a)$p=0.1625$ $(93\uparrow,67\downarrow)$\,,
(b)$p=0.475$ $(118\uparrow,42\downarrow)$\,, and (c) $p=0.9375$ $(155\uparrow,5\downarrow)$\,. 
All other parameters are the same as in Fig.\ref{susceptibility} (a).
 }
\label{Mstructure}
\end{figure}
We further increase the population imbalance ratio to examine the
spin-density distribution profiles before and after $p_{c}$. 
As for $p=0.1625$ $(N_\uparrow-N_\downarrow=26)$\,, 
as seen in Fig.\ref{Mstructure}(a)\,, the central Mott core is not magnetized, 
and all the polarization concentrates on the ferromagnetic metallic phases, 
while the magnetization amplitude is considerably enhanced compared to Fig.\ref{edge_spin1/2}. 
A coupling with the Haldane phase and polarized ferromagnetic metal brings 
about the enhancement of the edge magnetization. 
On the other hand, when $p=0.475 (> p_{\rm c})$, the staggered structure disappears, and
even the $p$-band Mott core is uniformly magnetized, as shown in
Fig.\ref{Mstructure}(b).  
As for an extremely population imbalance ratio $p=0.9375$, a periodical
oscillation appears inside the $p$-band Mott core as shown in
Fig.\ref{Mstructure}(c).   
This oscillation can be explained as follows. 
A low-energy excitation of the spin-$S$ Heisenberg model under strong
magnetic field is described by the Luttinger liquid irrespective of
the spin length $S$\,\cite{LL}. 
Then, this theoretical picture predicts that the periodicity of a
spin-density wave (SDW) is characterized by $2k_{\rm F}$, where 
$k_{\rm F}$ is the Fermi wave vector in the equivalent spinless fermion
system. 
In the present trapped systems, the oscillation periodicity is given by
the relation 
\(
2k_{\rm F}=(2\pi/L_{\rm Mott}) N_{{\rm Mott},\downarrow}
\)\,\cite{LL,Machida}, where $L_{\rm Mott}$ is the size of the 
Mott core, and $N_{{\rm Mott},\downarrow}$ is the total number of
spin-down component inside the Mott-core. 
Indeed, the magnetization profile shows an expected oscillation. 
In Fig.\,\ref{Mstructure}(c), the number of the minimum peaks of the
SDW is $5$. 
This number should be almost equal to $N_{{\rm Mott},\downarrow}$. 
In fact, we can directly evaluate $N_{{\rm Nott},\downarrow}$ and find
$N_{{\rm Mott},\downarrow}\simeq 5$. 
A more systematic analysis on $p$-dependence in such a
SDW-like oscillation is shown in Ref.\,\cite{supplementary}. 
In conclusion, we confirmed that the Haldane phase emerges as a leading
phase when utilizing 2nd degenerate $p$-orbitals in 1-D FGOL. 
At the half-filling condition for $p$-orbitals in the large $U$ limit,
the multi-band Hubbard Hamiltonian was reduced to the $S=1$ Heisenberg 
chain model. 
Then, the emergence of the Haldane phase and its associated physics were demonstrated by
DMRG studies on the original model together with the harmonic trap
potential. 
The polarization on the outer regions is easily observed in experiment\,\cite{phase separation1,phase separation2} 
and spin-selective single-site addressing\,\cite{singlespin} can be used to detect 
the staggered magnetization on Mott core. 
The formation of the Haldane phase in 1-D FGOL allows us to not only
investigate its features but also open a new avenue towards
topologically-protected quantum state engineering\,\cite{Miyake1}. 
 
We wish to thank T.~Morimae, H.~Onishi and R.~Igarashi for their useful
discussions. 
This work was partially supported by the Strategic Programs for
Innovative Research, MEXT, and the Computational Materials
Science Initiative, Japan. 
The numerical work was partially performed on Fujitsu BX900 in JAEA. 
We acknowledge their supports from CCSE staff.

\appendix



\begin{thebibliography}{99}
\bibitem{condition}
We take the number of states kept $m_{\rm s}=400$ in all DMRG
calculations on the model. 
The maximum truncation error is order of $10^{-7}$ and 
$m_{\rm s}$-dependence of ground-state energy $E(m_{\rm s})$ is
$[E(400)-E(350)]/Lt \simeq 10^{-7}$.
\end{thebibliography}
\end{document}